\documentclass{aastex62}
\usepackage{amsmath,mathtools,amssymb,graphicx,float,braket,slashed}
\usepackage[caption=false]{subfig}
\usepackage{tensor,enumerate,color}
\hypersetup{
    linkcolor=blue,
    filecolor=blue,      
    urlcolor=blue,
    citecolor=blue
}
\def\be {\begin{equation}}
\def\ee {\end{equation}}
\def\bea {\begin{eqnarray}}
\def\eea {\end{eqnarray}}
\def\bc {\begin{center}}
\def\ec {\end{center}}
\def\nn {\nonumber}

\newcommand{\bet}{\beta}
\newcommand{\del}{\delta}
\newcommand{\gm}{\gamma}
\newcommand{\om}{\omega}
\newcommand{\sg}{\sigma}
\newcommand{\Gm}{\Gamma}
\newcommand{\Sg}{\Sigma}
\newcommand{\Om}{\Omega}
\newcommand{\Lm}{\Lambda}
\newcommand{\ov}{\overline}
\newcommand{\rls}{\rightleftharpoons}
\newcommand{\p}{\prime}
\newcommand{\pp}{\prime\prime}
\newcommand{\ppp}{\prime\prime\prime}
\newcommand{\wt}{\widetilde}
\newcommand{\epz}{\epsilon_0}
\newcommand{\ep}{\epsilon}
\newcommand{\vx}{{\bf{x}}}
\newcommand{\vk}{{\bf{k}}}
\newcommand{\vp}{{\bf{p}}}

\submitjournal{ApJ}
\begin{document}

\title{\textbf{Saha ionization equation in the early Universe }}

\correspondingauthor{S. Mallik}
\email{samir.mallik@saha.ac.in}

\author{Aritra Das}
\affil{HENPP Division, Saha Institute of Nuclear Physics, HBNI, 
        1/AF Bidhan Nagar, Kolkata 700064, India.} 
\email{aritra.das@saha.ac.in}

\author{Ritesh Ghosh}
\affil{Theory Division, Saha Institute of Nuclear Physics, HBNI, 
        1/AF Bidhan Nagar, Kolkata 700064, India.}
\email{ritesh.ghosh@saha.ac.in}

\author{S. Mallik,}
\affil{Theory Division, Saha Institute of Nuclear Physics,
        1/AF Bidhan Nagar, Kolkata 700064, India.}
\email{samir.mallik@saha.ac.in}

\begin{abstract}
The Saha equation follows from thermal equilibrium of matter and radiation.
We discuss this problem of equilibrium in the early Universe, when matter
consists mostly of electrons, protons and hydrogen atoms. Taking H-atoms in
their ground state only and applying the real time formulation of thermal
field theory, we calculate the difference of ionization and recombination 
rates, which controls the equilibration of H-atoms. By comparing with
realistic calculations including the excited states of H-atom, we conclude
that the presence of excited states lower the equilibrium temperature from 5000 K to
4000 K. 
\end{abstract}

\section{Introduction} \label{introduction}
A century ago Saha \citep{Saha} used the thermodynamics of chemical equilibrium
to find the degree of ionization of atoms in a thermal bath. Since then it
has been vigorously applied to investigate the spectrum of the sun and other 
stars. More recently, it has been used to find the epoch of hydrogen 
recombination involving the reaction
\be
\gm \quad+\quad  H \quad\rls\quad p^+\quad +\quad e^-  ,
\label{int}
\ee
in the early Universe, leading to an understanding of the cosmic microwave 
background (CMB) radiation observed today \citep{Penzias,Dicke}. (There was an earlier epoch of Helium recombination, which can be treated separately in a first approximation.)

In the early Universe the equilibrium condition is not guaranteed a priori; it
depends on the reaction rate and cosmic expansion rate. Also the Saha equation 
neglects the excited states of the H-atom, which is a very good approximation 
as long as thermal equilibrium prevails. But away from equilibrium the excited 
states may be important in the process. Accordingly a number of authors 
\citep{Peebles,Zeldovich} have investigated the hydrogen ionization and 
recombination in the realistic case, without assuming equlibrium 
and including the excited states (2s, 2p) of the H-atom along with the ground (1s) state. The discovery of CMB anisotropies prompted a resurge of the calculation of recombination including non-leading effects. Thus the earlier (effective) three level calculation was replaced with a multi-level one \citep{Seager} including also Helium and their higher excited states. The effect of Raman scattering along with the related two-photon emission \citep{Chluba} was incorporated in the calculation. A review of all such effects is contained in the Karl Schwarzschild lecture by Sunyaev and Chluba \citep{Sunyaev}.

In this note, we attempt an indirect but easy way to study the effect of the
excited states in attaining equilibrium. We calculate the ionization
and recombination rates in a simplified model, where we exclude the excited 
states, taking only the H-atom in the ground state. Comparing these reaction
rates with the cosmic expansion rate, we may know at which
temperature the equilibrium is lost in the simplified model. On the other hand, 
comparing the fractional hydrogen ionization from the Saha equation with that of 
the realistic calculation (including the excited states), we can find when 
equilibrium is lost in the real world. Now comparing the two results for the 
loss of equilibrium, we may see the role of the exited states of H-atom. 

We use the reaction rates to write 
the Boltzmann equation for an arbitrary (nonequilibrium) distribution of
H-atoms. It has a simple analytic solution consisting of two terms
\citep{Weldon}. The first term gives the equilibrium distribution, while the 
second term vanishes exponentially with time. It is the coefficient of time 
in the exponential of this term, which is identified with the reaction 
rate tending the distribution to equilibrium. This rate must be compared with 
the expansion rate of the Universe, given by the-then Hubble parameter. 

Though the problem is non-relativistic, we shall use relativistic
expressions to evaluate the rates and then apply non-relativistic approximation. 
We use units with $\hslash(=h/2\pi)$ and $c$ taken to be unity, where $h$ and 
$c$ are Planck's constant and the velocity of light respectively. Also we write 
$\beta=1/k_B T$, where $k_B$ is the Boltzmann constant and $T$ the temperature.  
 
\section{Ionization and recombination rates}
In atomic physics the processes represented by \eqref{int} are well-known.
Recombination (or recapture) is the capture of an electron in the continuum
by the atomic nucleus with the emission of photon. Ionization (or photoeffect)
is the inverse process, where a photon is absorbed by an atom accompanied
by ejection of an electron. These transitions are caused by the electric dipole
operator in quantum mechanics \citep{Bethe,WeinbergQ}. It is given by the potential function, 
$V=-(e/m)\bf{A}\cdot\bf{p}$, where $e, m$ and $\bf{p}$ are the electric 
charge, mass of the electron and its momentum operator and $\bf{A}$ is the 
radiation field. The transition amplitudes for the above 
processes are given by the matrix elements of $V$ between two atomic states, 
where one is discrete and the other in the continuum.

Instead of using the above method to calculate the reaction rates, we use 
the elegant method of thermal quantum field theory for the problem 
~\citep{Semenoff,Niemi,Mallik}, where we directly get the recombination and 
ionization probabilities multiplied 
by appropriate factors involving distribution functions for particles in the 
medium. We first construct the interaction Lagrangian 
involving all the particles in \eqref{int}. Let the photon, electron and proton fields
be respectively $A_\mu (x)$, $\psi_e (x)$ and $\psi_p (x)$. We take H-atoms in 
the ground state only (ignoring its excited states), when it can be represented by 
an elementary scalar field $\Phi (x)$ \citep{WeinbergF,Dashen}. Then the 
required effective interaction Lagrangian is
\be
\mathcal{L}_{\mbox{int}}=g(\Phi\ov{\psi}_p\gm^\mu\psi_e+ h.c.)A_\mu \,.
\label{lint}
\ee
It describes an electromagnetic interaction and $g$ must have mass dimension
$-1$. So we take
\be
g=\frac{e}{m}\,, \label{g}
\ee
to within some uncertainty.

\begin{figure}[tbh]
\begin{center}
\includegraphics[scale=.35]{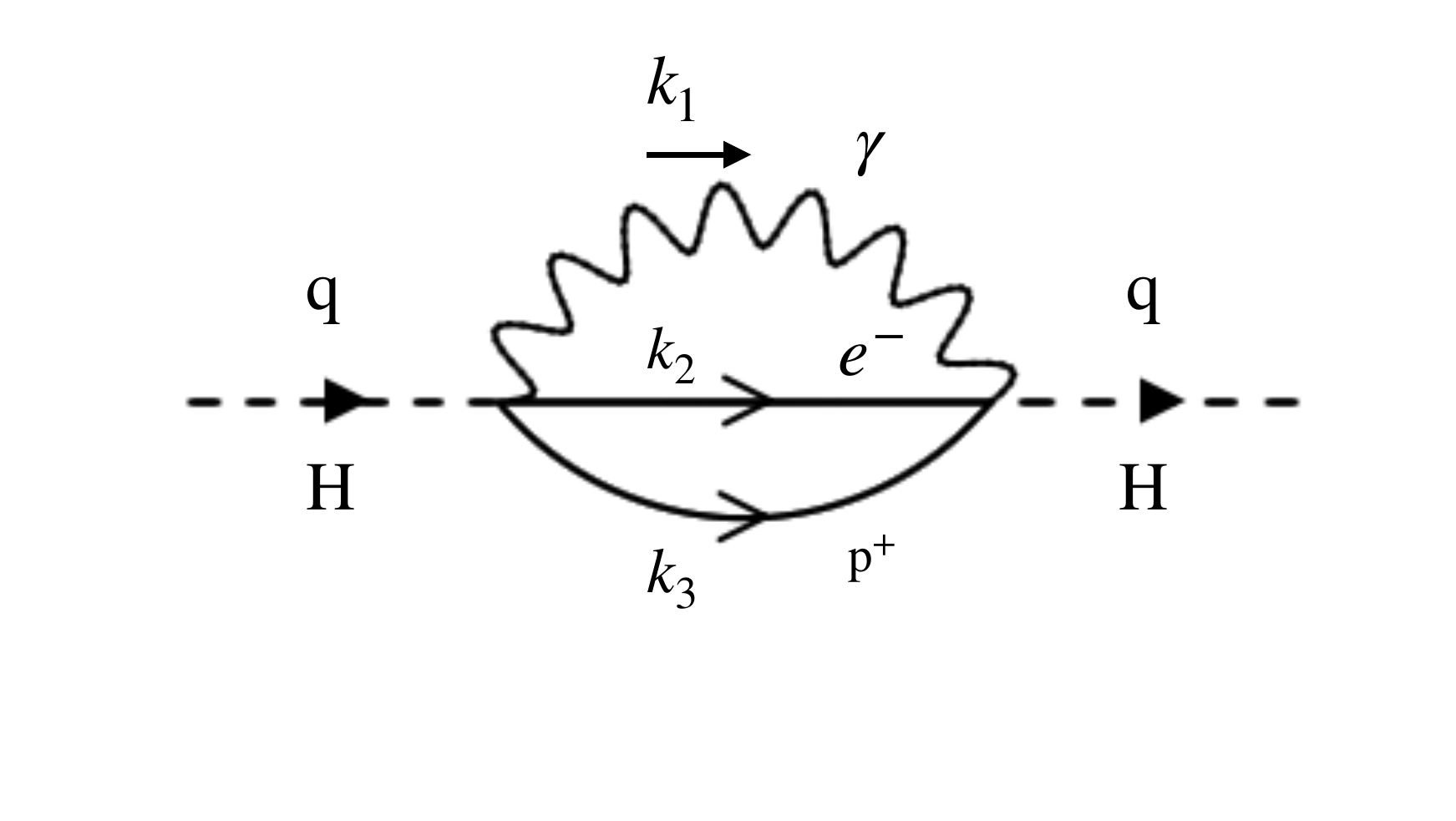}
\caption{A self-energy graph for H-propagator}
\label{feynman}
\end{center}
\end{figure} 
The calculation of the relevant reaction rate in thermal field theory can be
best approached by considering the self-energy graph for the H-atom shown in 
Fig.~\ref{feynman}. It includes in the intermediate state all other particles 
which appear in \eqref{int}. The elements of real-time thermal field theory is 
sketched in Appendix \ref{app1}. These elements are used in Appendix \ref{app2} 
to calculate this graph, from which we obtain the different imaginary parts, 
which are collected in (\ref{final}) in a compact form. Any particular imaginary 
part may be isolated by integrating the variables $k_0'$, $k_0''$ and $k_0'''$ 
in (\ref{final}) over the appropriate delta functions in the spectral functions, 
such as in \eqref{spec_boson}. 

\begin{center}
\begin{figure}[tbh]
\begin{center}
\includegraphics[scale=.4]{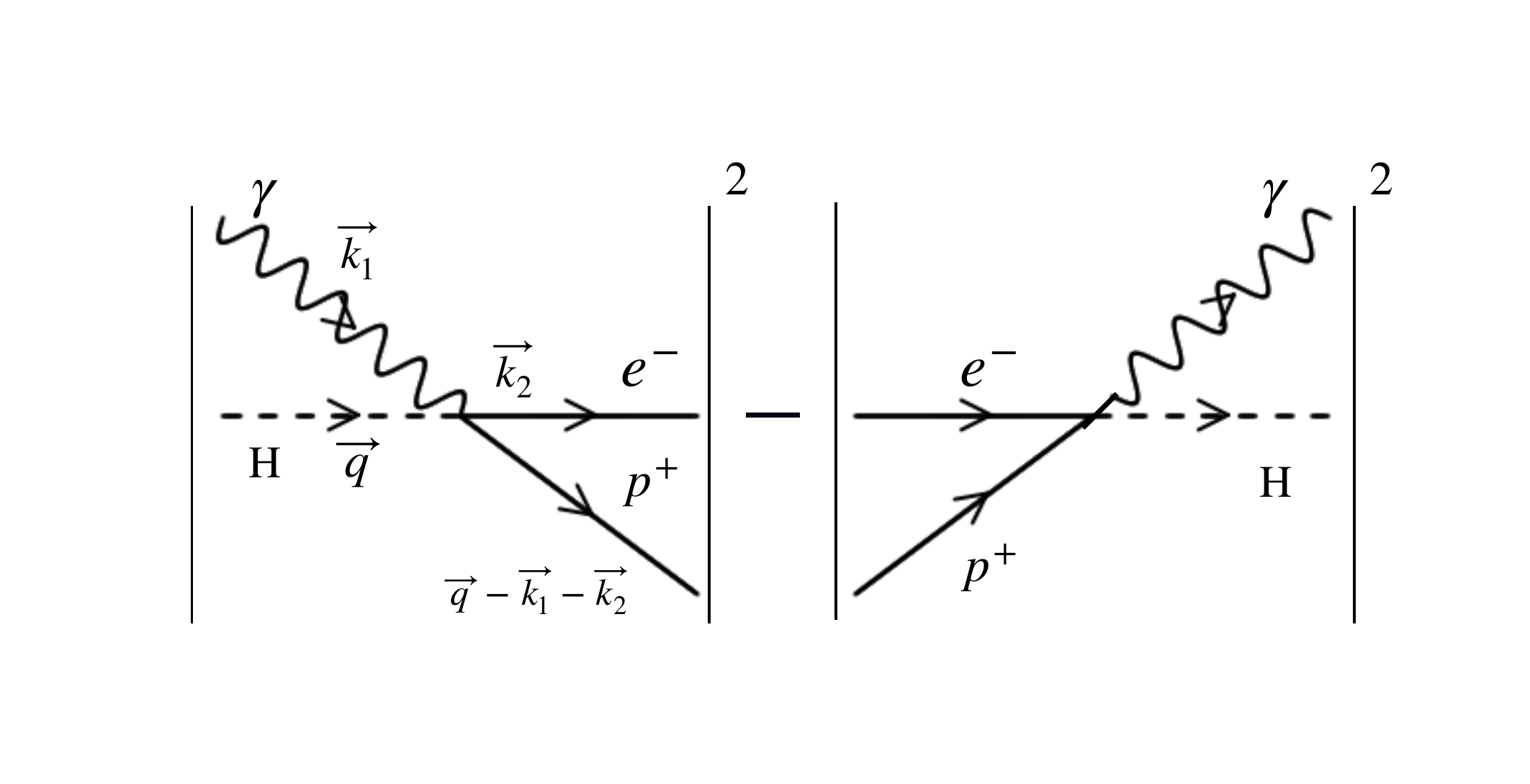}
\caption{The relevant imaginary part of Fig \ref{feynman}}
\label{imaginary}
\end{center}
\end{figure}
\end{center}

Before we write the desired imaginary part, we establish our notation. We
shall not use the four-momentum notation anymore; instead, we now denote the
magnitudes of three-momenta of H-atom (of mass $M_H$), photon, electron (of
mass $m$) and proton (of mass $M$) by $q,~ k_i~ (i=1,2,3)$ and energies by
$\om=\sqrt{q^2+M_H^2},~\om_1=k_1,~\om_2=\sqrt{k_2^2+m^2}$ and
$\om_3=\sqrt{k_3^2+M^2}$ respectively. 
We are interested in the imaginary part shown in Fig \ref{imaginary},
corresponding to processes in \eqref{int} with the photon (and H-atom) incoming 
and electron and proton outgoing for ionization, when the reverse process 
(recombination) will automatically be given by the second term in bracket in  \eqref{final}.
Accordingly we choose the delta functions in $k_0',~k_0''$ and $k_0'''$ 
variables as $\del(k_0'+\om_1)$, $\del(k_0''-\om_2)$ and $\del(k_0'''-\om_3)$. 
On using (A6) and (A11) to convert $f$ and $\wt{f}$ to $n$ and $\wt{n}$, we
get for the required processes (denoted by subscript 1) from \eqref{final},
\be
\mathrm{Im}\ov{\Sg}_{(1)}=16\pi g^2mM\int 
\frac{d^3k_1}{(2 \pi)^32\om_1}\frac{d^3 k_2}{(2 \pi)^32\om_2}\frac{1}{2 \om_3}
[n_1(1-\wt{n}_2)(1-\wt{n}_3)-(1+n_1)\wt{n}_2\wt{n}_3]
\del(\om+\om_1-\om_2-\om_3)\,,
\label{s}
\ee
where $n_1(\om_1),~\wt{n}_2(\om_2)$ and $\wt{n}_3(\om_3)$ are the equilibrium 
distribution functions for the photon, electron and proton respectively.
It resembles the unitarity relation for the S-matrix in vacuum for the
two-particle states of photon and H-atom. Compared to that relation, we now
have the difference of two terms. Dividing it by $\om$, we convert it to rates 
\citep{Weldon}
\begin{align}
\frac{\mathrm{Im}\ov{\Sg}_{(1)}}{\om}=\Gm_d-\Gm_i \,,
\end{align}
where $\Gm_d$ and $\Gm_i$ are the first and the second term in (4), representing 
respectively the decay and inverse decay rates of H-atom.

\section{Boltzmann equation}
The rates $\Gm_d$ and $\Gm_i$ are related. To see this, we write the
equilibrium distribution functions explicitly as 
\be
n_1=\frac{\exp\left(-\bet \om_1/2\right)}{\exp(\bet \om_1/2)-\exp(-\bet
\om_1/2)}, \qquad 1+n_1=\frac{\exp\left(\bet \om_1/2\right)}
{\exp(\bet \om_1/2)-\exp(-\bet \om_1/2)}\nn
\ee
and similarly for factors involving $\wt{n}_2$ and $\wt{n}_3$. Taking into 
account the energy conserving delta function in (\ref{s}), we get
\be
\Gm_d\,\,;\, \Gm_i =16\pi g^2m\frac{M}{\om} \{\exp[\bet(\om-\mu_2-\mu_3)/2]\,\,;\, 
\exp[-\bet(\om-\mu_2-\mu_3)/2]\}\times L \,, \label{gdbgi}
\ee
where $\mu_2$ and $\mu_3$ are the chemical potentials for the electron and
proton and $L$ is given by 
\be
L=\int\frac{d^3k_1}{(2\pi)^32\om_1}\frac{d^3k_2}{(2\pi)^32\om_2}\frac{1}{2\om_3}
\frac{\del(\om+\om_1-\om_2-\om_3)}{\displaystyle \prod\left\lbrace\exp[\bet
(\om_i-\mu_i)/2]\mp \exp[-\bet (\om_i-\mu_i)/2]\right\rbrace} \,,\label{L}
\ee
where the product runs over $i=1,2,3$. For $i=1,~ \mu_1=0$ and we take the upper
sign and for $i =2,3$ we take the lower sign. So we get the ratio
\be
\frac{\Gm_d}{\Gm_i}=\exp[\bet(\om-\mu_2 -\mu_3)] \,. \label{d_by_i}
\ee

So far we have treated the H-atom as a single particle without any distribution 
in the medium. Let us now assume an arbitrary (non-equilibrium) 
distribution  $n(\om,t)$ of these particles. We can write a Boltzmann equation 
for $n(\om,t)$, noting that it decreases at the rate $n\Gm_d$ and increases 
at the rate $(1+n)\Gm_i$ ,
\be
\frac{dn(\om,t)}{dt}=(1+n)\Gm_i-n \Gm_d \,,
\ee
whose solution is \citep{Weldon,Bellac}
\begin{align}
n(\om,t)&=\frac{\Gm_i}{\Gm_d-\Gm_i}+c(\om)e^{-(\Gm_d-\Gm_i)t}\nn\\
&=\frac{1}{\exp{\bet(\om-\mu_2 -\mu_3)}-1}+c(\om)e^{-\Gm t},\quad \Gm=\Gm_d-\Gm_i\,,
\label{sol_bol}
\end{align}
where $c(\om)$ is an arbitrary function and we use (8). If $\mu$ is the
chemical potential of the H-atom, the first term in $n(\om,t)$ satisfies the 
condition 
\be
\mu=\mu_2+\mu_3, 
\ee
which is the condition of chemical equilibrium \citep{Reif}, as can be read off 
from (1). Observe that this condition arises automatically in our
calculation as a result of using the equilibrium thermal propagators. So the
distribution function approaches the equilibrium value exponentially in time, 
irrespective of its initial distribution and its rate is governed by the
reaction rate $\Gm$.

So far the formulae are exact. Let us now make two simplifications appropriate 
for the problem at hand. First, we write the energies in nonrelativistic
approximation:
\be
\om=M_H+\frac{q^2}{2M_H},\qquad \om_1=k_1, \qquad \om_2=m+\frac{k_2^2}{2m},
\qquad \om_3=M+\frac{k_3^2}{2M} 
\ee 
with $M_H=M+m-\epz$, where $\epz (=13.6~ eV)$ is the binding energy of the
H-atom in the ground state, which we neglect except in the exponential. Our
second smplification results from the particle densities being dilute. We
replace Fermi-Dirac and Bose-Einstein distributions with those of
Maxwll-Boltzmann:
\be
\wt{n}_2=\exp\left[-\bet\left(\frac{k_2^2}{2m}-\mu'_2\right)\right], 
\qquad \wt{n}_3=\exp\left[-\bet\left(\frac{k_3^2}{2M}-\mu'_3\right)
\right], \qquad
n=\exp\left[-\bet\left(\frac{q^2}{2M}-\epz-\mu'_2-\mu'_3\right)
\right]\,,
\ee
where we define non-relativistic chemical potentials by $\mu'_2=\mu_2-m$ and 
$\mu'_3=\mu_3-M$. 

The total number of electrons in volume $V$ is
\be
N_e=2V\int\frac{d^3k}{(2\pi)^3}\wt{n}_2(k)=2V\left(\frac{m}{2\pi\bet}
\right)^{3/2}\exp(\bet\mu'_2)\,. \label{N_e}
\ee
Similarly, the total number of protons and H-atoms are
\begin{align}
N_p &=2V\left(\frac{M}{2\pi\bet}\right)^{3/2}\exp(\bet\mu'_3) \,,
\label{N_p}\\
N_H &=4V\left(\frac{M_H}{2\pi\bet}\right)^{3/2}\exp[\bet(\epz+\mu'_2+\mu'_3)] \,,
\label{N_H}
\end{align}
which incorporates the equilibrium condition (11).
We then get the Saha equation \citep{WeinbergC}
\be
\frac{N_e N_p}{N_H}= V\left(\frac{ m}{2\pi\bet}\right)^{3/2}\exp(-\bet\epz)\,.
\ee

\section{early Universe}
We now examine the equilibrium condition in the early Universe. We first 
estimate the reaction rate $\Gm$. As in the previous Section, we reduce the 
distributions to that of Maxwell-Boltzmann by retaining  only the positive 
exponentials in the products in the denominator of \eqref{L}. Also we ignore 
terms of $O(1/M)$ compared to $O(1/m)$ and write $\om_2\om_3\simeq mM$ in this 
denominator to get $L$ as
\begin{align}
L=\left(\frac{4\pi}{(2\pi)^3}\right)^2\frac{1}{8Mm}\int_0^\infty
dk_1\,k_1\,dk_2\,k_2^2\,\exp\left[-
\frac{\bet}{2}\left(k_1+\frac{k_2^2}{2m}-\mu'_2
-\mu'_3\right)\right]\del\left(-\epz+k_1-\frac{k_2^2}{2m}\right)\,.
\end{align}
We remove the $k_1$ integral with the delta function, when the $k_2$
integral reduces to a Gamma function, giving
\be
L=\left(\frac{4\pi}{(2\pi)^3}\right)^2\frac{1}{8Mm}\exp[-\bet
(\epz-\mu'_2-\mu'_3)/2]\frac{\sqrt{\pi}}{4}
\left(\frac{m}{\bet}\right)^{3/2}\left(\epz+\frac{3}{4\bet}\right)\,. \label{Lf}
\ee
As the H-atom concentration is dilute, we set $1+n\simeq 1$ in (9), when the
solution (10) shows the equilibrium distribution to be of Maxwell-Boltzmann 
type and the reaction rate becomes $\Gm\simeq \Gm_d$. From (6) and (19)
we thus get
\be
\Gm=\frac{g^2\sqrt{\pi}}{8\pi^3M}\left(\frac{m}{\bet}\right)^{3/2}\epz
\exp(-\beta \epz)\left(1+\frac{3}{4\bet\epz}\right) \,.
\ee
Noting (3) we may rewrite it as
\be
\Gm\simeq \frac{e^2}{4\pi}\frac{1}{2\pi^{3/2}}\frac{m}{M}
\left(\frac{k_BT}{m}\right)^{3/2}\epz\exp(-\epz/{k_BT})\,.
\ee

Next, the expansion rate of the Universe is given by the Hubble pamameter 
$H=\dot{a}/a$, where $a(t)$ is the scale factor in the metric. In the era of
interest to us ($T>1000 K$), the constant vacuum energy is utterly negligible
and we consider the energy density of matter, 
both non-relativistic ($\rho_M$) and relativistic ($\rho_R$). Also we do not 
include the curvature term, which, assuming a prior inflationary epoch, is
driven to $0$.
Denoting the present values by the subscript (0), we then write the total energy
density as
\be 
\rho(t)= \rho_0\left[ \Om_M\left(\frac{T}{T_0}\right)^3 +
\Om_R\left(\frac{T}{T_0}\right)^4 \right]\,,
\ee
where $\Om_M$ and $\Om_R$ are fractions of the present critical energy density
$\rho_0$. Applying Einstein equation with the spatially flat metric, it gives
\be
H^2 =H_0^2\left[ \Om_M\left(\frac{T}{T_0}\right)^3 +
\Om_R\left(\frac{T}{T_0}\right)^4 \right]\,.
\ee
Including the contributions of photons and neutrinos in $\Om_R$ and putting
in numbers, one gets \citep{WeinbergC}
\be
H=7.20\times 10^{-19}T^{3/2}\left(\Om_Mh^2 +1.52\times10^{-5}T\right)^{1/2} 
s^{-1}\,,
\ee
where $h$ is the Hubble constant in units of 100km/sec/Mpc. We take $\Om_M
h^2=0.15$ as in Ref.\citep{WeinbergC}.

In Table 1, we show the reaction rate ($\Gm$) and the expansion 
rate ($H$) of the Universe at temperatures in the region of interest. It is seen 
that equilibrium prevails up to about 5000 K, if we include only the ground 
state of H-atom in calculating the reaction rate.

\begin{deluxetable}{ ccc }
\tablecaption{Reaction rate ($\Gm$) and expansion rate ($H$) at different 
temperatures} \label{tab:title}
\tablehead{
\colhead{$T$(in K)} & \colhead{$\Gm$ in $\mbox{s}^{-1}$} & \colhead{$H$ in 
$\mbox{s}^{-1}$}
}
\startdata
6000 & \hspace{.1cm} 2.6$\times 10^{-11}$ \hspace{2.5cm} & 1.6$\times 10^{-13}$ \\
5000 & 1.0$\times 10^{-13}$ & 1.2$\times 10^{-13}$ \\
4000 & 2.7$\times 10^{-17}$ & 8.4$\times 10^{-14}$ \\
3000 & 3.4$\times 10^{-23}$ & 5.2$\times 10^{-14}$ \\
\enddata
\end{deluxetable}
\section{result and conclusion}
In this work we do not calculate the fractional hydrogen ionization. We only
estimate the role of excited states of H-atom in attaining the equilibrium
condition as the temperature falls in the early Universe. We have already estimated above the temparature up to which equilibrium condition prevails in the case of our simplified model with no excited states of H-atom. We now get its estimate from the complete calculation of fractional ionization including the excited states and other physical effects presented in figure 3 of Sunyaev and Chluba \citep{Sunyaev}. Because the Saha equation assumes equilibrium condition, this condition should prevail as long as the complete calculation agrees with the Saha equation. We see that the two calculations start to disagree at redshift z $\sim 1500$ corresponding to T $\sim 4000$ K. Comparing this result with our calculation, we see that the excited states bring down the equilibrium temperature from $5000$ K to $4000$ K 

Thus although the equilibrium number density of excited H-atoms 
is negligible compared to that in the ground state, they provide pathways to 
facilitate attaining the equilibrium and our calculation gives a quantitative 
estimate of this effect. Next, for results at low ionization levels, in which 
the last photon scattering took place, there is appreciable deviation from 
equilibrium \citep{Peebles}. Still, the Saha equation gives an order of magnitude 
estimate of the recombination temperature in the early Universe \citep{Kolb}. 
 
Finally we comment on the use of (real time) thermal field theory. The
factors involving the distribution functions in reaction probabilities in a
medium are known, since Einstein introduced the $A$ and $B$ coefficients by 
considering detailed balance of equilibrium of atoms in the radiation field~ 
\citep{Einstein}. As is well-known, they appear in matrix elements of
creation and destruction operators in quantum field theory. These factors
are now put in by hand in all processes taking place in a
medium~\citep{Uehling33,WeinbergP79}. Here we show that they arise naturally 
originating from the thermal propagators. Another advantage of
thermal field theory is that the polarization sums are done automatically in
reaction probabilities.
\section{Acknowledgement}
One of us (S.M.) wishes to thank Prof. T. Souradeep for sending him his
article entitled `Meghnad Saha and the Cosmic Photosphere'. He also thanks
Prof. R. Rajaraman for a discussion.
 
\appendix

\section{Equilibrium thermal field theory} \label{app1}
Here we recall the basic elements of equilibrium thermal field theory 
\citep{Mallik}. Compared to vacuum field theory, it differs essentially in the 
time path and hence the propagator. To bring out this difference let us take 
a scalar field $\phi(x)$, with $x^{\mu}=(\tau,\vx)$, where the time variable 
$\tau$ may be complex and consider the time ordered propagator, which arises 
in perturbative calculations. While for the vacuum propagator 
\be
\Delta_F(x,x')=i\braket{0|T\phi(t,\vx)\phi(t',\vx')|0},
\ee
the time variables run over the real time axis, the corresponding thermal 
propagator
\begin{align}
D(x,x')=i\frac{\mbox{Tr}[\exp(-\bet H)\phi(t,\vx)\phi(t',\vx')]}{\mbox{Tr}
\exp(-\bet H)}
\end{align}
has the time variables running over an interval in the complex time plane. 
Here $H$ is the hamiltonian of the system and the thermal trace is over a
complete set of states. 

\begin{center}
\begin{figure}[tbh]
\begin{center}
\includegraphics[scale=0.34]{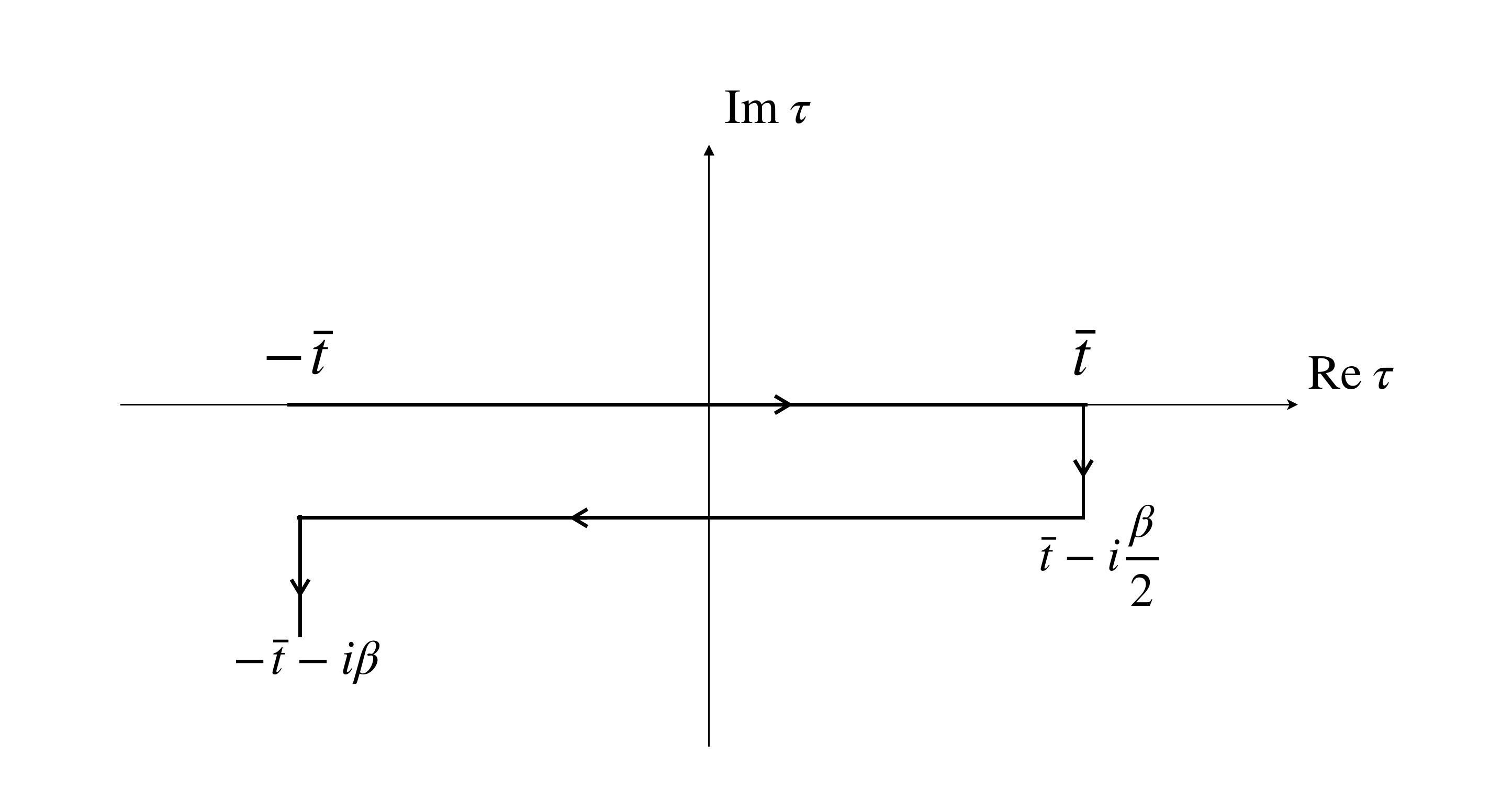}
\caption{The time contour in the real time formulation of thermal field
theory with $\bar{t}\longrightarrow \infty$.}
\label{time}
\end{center}
\end{figure}
\end{center}

The time path at finite temperature may be broadly chosen in two different ways. 
It may be the imaginary segment from 0 to $-i \bet$ in the complex $\tau$
plane, giving the imaginary time formulation. In the real time formulation, 
which we shall use here, the time path must traverse the real axis. Then it must end at a point with Im $\tau=-\bet$. 
There are different ways to complete this path; we shall choose the one shown 
in Fig~\ref{time}. Only the two horizontal lines contribute to the propagator, 
making the propagator a $2\times 2$ matrix.

It is possible to write spectral representation for propagator of fields of 
any spin. For the problem at hand, we have the scalar field $\Phi(x)$ to 
represent the H-atom, the vector field $A_\mu(x)$ for the photon and the Dirac 
field $\psi_e(x)$ and $\psi_p(x)$ for the electron and proton. The form of the 
spectral representation depends on the bosonic or fermionic nature of the field.
The $2\times 2$ matrix propagator for the scalar field is 
\be
D_{ij}(k_0,\vk)=\int_{-\infty}^{+\infty}\frac{d k_0'}{2 \pi}\rho(k_0',\vk)
\Lm_{ij}(k_0',k_0) , \qquad (i,j=1,2) \label{prop} \nn
\ee
where the spectral function is
\begin{align}
\rho(k_0',\vk)&=2\pi\ep(k_0')\delta(k_0'^2-\om^2) \nn \\
&=2\pi\ep(k_0')\frac{1}{2\om}\{\delta(k_0'-\om)+
\delta(k_0'+\om)\}, \qquad \om=\sqrt{\vk^2+m^2} \,. \label{spec_boson}
\end{align}
Here $\ep(k_0)$ is the sign function defined as $+1$ for $k_0>0$ and 
$-1$ for $k_0<0$. As we shall see, only the $11$-component of the propagator 
appears in our calculation, for which
\be
\Lm_{11}(k'_0,k_0)=\frac{1+f(k'_0)}{k_0'-k_0-i\eta}-\frac{f(k'_0)}{k_0'-k_0+i\eta}\,,
\ee
where $f(k_0')$ is a distribution-like function
\begin{align}
f(k'_0)=\frac{1}{\exp[\bet(k_0'-\mu)]-1}\,.
\end{align}
We shall use the $\del$-function in the spectral function to express it 
in terms of the true distribution function $n(\om)$,
\begin{align}
f(\om)=\frac{1}{\exp[\bet(\om-\mu)]-1}\equiv n(\om)\hspace{.1cm}, \hspace{.5cm} 
f(-\om)=-(1+n(\om))\,.
\end{align}
For the vector field $A_\mu(x)$, the propagator is again given by the one for the scalar propagator with the spectral function,
\begin{align}
\rho_{\mu\nu}(k'_0,\vk)=2 \pi \ep(k'_0) g_{\mu\nu} \delta(k'^2)\,. \label{n0}
\end{align}

For the spin $\frac{1}{2}$ (fermion) propagator, we have a similar 
representation
\bea
S_{ij}(p_0,\vp)=\int \frac{d k_0^\prime}{2 \pi} \hspace{.1cm} 
\sg(p_0^\prime,\vp)\hspace{.1cm} \Om_{ij}(p_0',p_0), \hspace{.5cm} (i,j=1,2)\nn
\\
\label{prop2}
\eea
with the spectral function
\bea
\sg(p'_0,\vp)=2 \pi \ep(p'_0) (\slashed{p}+m) \del({p'}^2-m^2)\nn
\\
\eea
and the 11-component of the $\Om$-matrix is 
\bea
\Om_{11}(p_0',p_0)=\frac{1-\wt{f}(p_0')}{p'_0-p_0-i\eta}+\frac{\wt{f}(p'_0)}
{p'_0-p_0+i\eta}\,.
\eea
Again $\wt{f}(p_0')$ can be written in terms of the fermion distribution 
function as 
\be
\wt{f}(\om)=\frac{1}{\exp[\bet(\om-\mu)]+1}\equiv \wt{n}(\om) \,.
\ee
(Also $\wt{f}(-\om)=1-(\exp[\beta(\om+\mu)]+1)^{-1}$, but we shall not need
it.)

The above free propagators and the interacting (complete) ones and hence the 
self-energies may be diagonalised. Here we are primarily interested in (the 
imaginary part of) the self-energy of the $\Phi(x)$ field representing H-atom, 
which diagonalizes as 
\begin{align}
\Sigma(q)=U^{-1}(q)
\begin{pmatrix}
	\ov{\Sg}(q) & 0\\
	0 & -\ov{\Sg}^*(q)
\end{pmatrix}
U^{-1}(q)\,,
\label{matrix}
\end{align}
where the diagonalizing matrix is 
\begin{align}
U^{-1}(q)=
\begin{pmatrix}
	\sqrt{1+n} & -\sqrt{n}\\
	-\sqrt{n} & \sqrt{1+n}
\end{pmatrix}
;
\hspace{.5cm}
n=\frac{1}{\exp(\bet|q_0|)-1}\,.
\end{align}
From \eqref{matrix} we get 
\begin{align}
\mathrm{Im} \hspace{.1cm} \ov{\Sg}(q)=\ep(q_0)\tanh(\frac{\bet q_0}{2}) 
\hspace{.2cm} \mathrm{Im} \hspace{.1cm} \Sg_{11}(q)\,,
\label{im}
\end{align}
so that we may evaluate only the 11-component of the $\Sg$ matrix to get the 
imaginary part of the diagonalized matrix.

\section{Evaluation of self-energy graph} \label{app2}
Here we shall evaluate the two-loop thermal self-energy graph (Fig~\ref{feynman}) 
of the H-atom. As the vacuum and medium calculations differ only in the 
propagators, we can first conveniently find it in vacuum: 
\begin{align}
\Sg(q)&=-2 g^2\int  \frac{d^4 k_1}{(2 \pi)^4} \frac{d^4 k_2}{(2 \pi)^4} 
\frac{d^4 k_3}{(2 \pi)^4} (2\pi)^4 \del^4(q-k_1-k_2-k_3)D^{\mu\nu}(k_1)
\mbox{tr}[S^{(e)}(k_2)\gm_\nu S^{(p)}(k_3)\gm_\mu]\,,
\end{align}
where \emph{tr} indicates trace over $\gm$ matrices and $D^{\mu\nu}$, $S^{(e)}$ 
and $S^{(p)}$ are the vacuum propagators for the photon, electron and proton 
respectively. Then the 11-component of the thermal self-energy matrix is 
immediately obtained by replacing the vacuum propagators with the corresponding 
$11$-component of the thermal propagator matrices: 
\begin{align}
\Sg_{11}(q)=-2 g^2\int  \frac{d^4 k_1}{(2 \pi)^4} \frac{d^4 k_2}{(2 \pi)^4} 
\hspace{2mm}D_{11}^{\mu\nu}(k_1)
\mbox{tr}\left[S^{(e)}_{11}(k_2)\gm_{\nu} S^{(p)}_{11}(q-k_1-k_2)\gm_{\mu}\right]\,,
\label{sigma11}
\end{align}
where we keep only the independent loop momenta. We write the propagator in 
their spectral representations (\ref{prop}) and (\ref{prop2}). The tensor and 
spinor factors in the spectral functions can be collected to give 
\begin{align}
g^{\mu\nu}\mbox{tr}[(\slashed{k_2}+m)\gm_\nu
(\slashed{q}-\slashed{k_1}-\slashed{k_2})\gm_\mu]\simeq 8Mm\,.
\end{align}
Removing these factors, we get the three spectral functions as
\begin{align}
\rho(k_1) &= 2 \pi \ep(k_{10}) \del(k_1^2)\,, \nn \\
\sg_{e}(k_2) &= 2 \pi \ep(k_{20}) \del(k_2^2-m^2)\,, \nn \\
\sg_{p}(k_3) &= 2 \pi \ep(k_{30}) \del(k_3^2-M^2)\,,
\end{align}
where $k_3 = q -k_1 -k_2$. Also we segregate the integrals in energy components 
of $k_1$ and $k_2$ over the energy denominators of propagators. We thus write 
\eqref{sigma11} as
\begin{align}
\Sg_{11}=-16 g^2 mM \int \frac{d^3 k_1}{(2 \pi)^3} 
\frac{d^3 k_1}{(2 \pi)^3} \int \frac{d k_0^{\p}}{(2 \pi)}\frac{d k_0^{\pp}}
{(2 \pi)}\frac{d k_0^{\ppp}}{(2 \pi)}\rho(k_0^{\p},\vec{k_1}) \sg_{e}(k_0^{\pp}
,\vec{k_2}) \sigma_{p}(k_0^{\ppp},\vec{q}-\vec{k_1}-\vec{k_2})\cdot K\,,
\end{align}
where
\bea
K=&&\int \frac{d k_{10}}{2 \pi} \frac{d k_{20}}{2 \pi} \bigg(\frac{1+f'}
{k_0'-k_{10}-i\eta}-\frac{f'}{k_0'-k_{10}+i\eta}\bigg)
\bigg(\frac{1-\wt{f}''}{k_0''-k_{20}-i\eta}+\frac{\wt{f}''}{k_0''-k_{20}+i\eta}\bigg)\nn\\
&&\times \bigg(\frac{1-\wt{f}'''}{k_0'''-(q_0-k_{10}-k_{20})-i\eta }
+\frac{\wt{f}'''}{k_0'''-(q_0-k_{10}-k_{20})+i\eta }\bigg)
\label{k}
\eea
with $f'=f(k_0')$,  $\wt{f}''=\wt{f}(k_0'')$, $\wt{f}'''=\wt{f}(k_0''')$.
		
Let us work out the integral over $k_{10}$ first, noting that this variable
appears in the first and third factors in (\ref{k}). When these two factors 
are multiplied out, these result four terms. As the integral converges in
both the upper and lower half of the $k_0$ plane, we can evaluate it closing
the integration contour in either half. So only two of these terms, having poles 
in both the upper and lower halves of $k_{10}$ plane, can contribute to the 
integral. Thus we evaluate the $k_{10}$ integral in (\ref{k}) to get
\bea
\bigg(\frac{(1+f')(1-\wt{f}''')}{k_{20}+k_0'+k_0'''-q_0+i\eta}+\frac{f' \wt{f}
'''}{k_{20}+k_0'+k_0'''-q_0-i\eta}\bigg)\,.
\eea
Next carry out the $k_{20}$ integral in the same way over the second factor in 
(\ref{k}) and the one just obtained to get
\bea
K=-\bigg(\frac{(1+f^{\p})(1-\wt{f}^{\pp})(1-\wt{f}^{\ppp})}{k_0^{\p}+k_0^{\pp}
+k_0^{\ppp}-q_0-i\eta}-\frac{f^{\p} \wt{f}^{\pp}\wt{f}^{\ppp}}{k_0^{\p}+
k_0^{\pp}+k_0^{\ppp}-q_0+i\eta}\bigg)\,,
\eea
giving its imaginary part as
\begin{align}
\mbox{Im}\,K &=-\pi[(1+f')(1-\wt{f}'')(1-\wt{f}''')+f'\wt{f}''
\wt{f}''']\del(q_0 - k_0' - k_0''-k_0''') \nn \\
&=-\pi\coth\left(\frac{\beta q_0}{2}\right)\left[(1+f')(1-\wt{f}'')
(1-\wt{f}''')-f'\wt{f}''\wt{f}'''\right]\del(q_0 - k_0' 
- k_0''-k_0''')\,.
\end{align}
Comparing with (\ref{im}) we finally get the imaginary part of the diagonalised self-energy as
\begin{align}
\mathrm{Im}\ov{\Sg}(q)&=16 g^2 mM \pi \ep(q_0) \int \frac{d^3 k_1}{(2 \pi)^3} \frac{d^3 k_2}{(2 \pi)^3} \int \frac{d k_0^{\p}}{(2 \pi)}\frac{d k_0^{\pp}}{(2 \pi)}\frac{d k_0^{\ppp}}{(2 \pi)} \rho(k_0^{\p},\vec{k_1}) \sigma_{e}(k_0^{\pp},\vec{k_2}) \sigma_{p}(k_0^{\ppp},\vec{q}-\vec{k_1}-\vec{k_2}) \nn \\
&\hspace{5cm} \times\left[(1+f^{\p})(1-\wt{f}^{\pp})(1-\wt{f}^{\ppp})-f^{\p} \wt{f}^{\pp}\wt{f}^{\ppp}\right]\delta\left(q_0 - k_0^{\p} - k_0^{\pp}-k_0^{\ppp}\right)\,.
\label{final}
\end{align}
Integrating over $k'_0, k''_0$ and $k'''_0$ with the delta functions
contained in the three spectral functions, we get eight terms 
corresponding to different particles in the initial and final states.
In Section 2 we get one of these terms representing the ionization and
recombination probabilities.


\begin{thebibliography}{}
\bibitem[Bethe(2008)]{Bethe} Bethe, H. A and Salpeter, E. E. 2008 , Quantum mechanic
and two-electron atoms, Dover Pub. New York
\bibitem[Chluba \& Thomas (2011)]{Chluba} Chluba, J and Thomas R.M. 2011, Mon. Not. R. Astron. Soc. 412, 748
\bibitem[Dashen(1974)]{Dashen}  Dashen, R. F. and Rajaraman, R. 1974 Phys. Rev.10, 708
\bibitem[Dicke(1965)]{Dicke} Dicke, R. H., Peebles, P. J. E., Roll, P. G. and  Wilkinson, D.T. 1965, ApJ. 142, 414
\bibitem[Einstein(1917)]{Einstein} Einstein, A. 1917, Phys. Zeit. 18, 121
\bibitem[Kolb(1989)]{Kolb}  Kolb, E. W. and  Turner, M. S. 1990, The early Universe, Addison-Wesley
\bibitem[Le Bellac(2000)]{Bellac} Le Bellac, M. 2000,  Thermal field theory, Cambridge University Press
\bibitem[Mallik \& Sarkar (2016)]{Mallik} Mallik, S. and  Sarkar, S. 2016,  Hadrons at Finite Temperature, Cambridge University Press
\bibitem[Niemi(1984)]{Niemi} Niemi, A. J. and Semenoff, G. W. 1984 Ann. Phys. 152, 105 
\bibitem[Peebles(1968)]{Peebles} Peebles, P. J. E. 1968, ApJ, 153, 1 
\bibitem[Penzias \& Wilson (1965)]{Penzias} Penzias, A. A \& R.W. Wilson, 1965, ApJ, 142, 419
\bibitem[Reif(1985)]{Reif} Reif, F. 1985, Fundamentals of Statistical and Thermal Physics, McGraw-Hill Book Company
\bibitem[(1920)]{Saha} Saha, M. N. 1920,  Phil. Mag., series 6, 40, 472,
809
\bibitem[Seager et al (1999)]{Seager} Seager, S., Sasslov, D. D. and Scott, D. 1999 ApJ, 523 , L1
\bibitem[Semenoff(1983)]{Semenoff} Semenoff, G. W. and Umezawa, H. 1983 Nucl. Phys. B220, 196
\bibitem[(2018)]{Sunyaev} Sunyaev, R.A and Chluba J. 2009, AN 330, 657
\bibitem[Uehling(1933)]{Uehling33} Uehling, E. A. and  Uhlenbeck, G. E. 1933, Phys. Rev.,43, 552
\bibitem[Weinberg(2008)]{WeinbergC} Weinberg, S. 2008, Cosmology, Oxford University Press
\bibitem[Weinberg(2013)]{WeinbergQ} Weinberg, S. 2013, Lectures in Quantum mechanics, Cambridge University Press,
\bibitem[Weinberg(1995)]{WeinbergF} Weinberg, S. 1995,  The Quantum Theory of Fields, vol I, Cambridge University Press 
\bibitem[Weinberg(1979)]{WeinbergP79} Weinberg, S. 1979 Phys. Rev. Lett., 42, 850
\bibitem[Weldon(1983)]{Weldon} Weldon, H. A. 1983, Phys. Rev.D 28, 2007
\bibitem[Zeldovich \& Sunyaev (1969)]{Zeldovich} Zeldovich, Ya.B and Sunyaev R.A., 1969 Astrophysics and Space Science, Volume 4,  3, 301

\end{thebibliography}
\end{document}